\documentclass[twocolumn,showpacs,preprintnumbers,prd,superscriptaddress,nofootinbib]{revtex4-1}

\usepackage{amsmath}
\usepackage{amsfonts}
\usepackage{amssymb}
\usepackage{graphicx}
\usepackage{color}
\usepackage{hyperref}
\usepackage{booktabs}
\usepackage{hyperref}
\usepackage{cleveref}
\usepackage{color}
\usepackage{multirow}

\Crefname{equation}{Eq.}{Eqs.}
\Crefname{figure}{Fig.}{Figs.}
\Crefname{section}{Sec.}{Secs.}

\usepackage{etoolbox}
\makeatletter
\appto{\appendix}{%
  \@ifstar{\def\theequation@prefix{A.}}%
          {}%
}
\makeatother

\begin{document}

\title{Cosmographic analysis with Chebyshev polynomials}

\author{Salvatore Capozziello}
\email{capozzie@na.infn.it}
\affiliation{Dipartimento di Fisica, Universit\`a di Napoli  ``Federico II'', Via Cinthia, I-80126, Napoli, Italy.}
\affiliation{Istituto Nazionale di Fisica Nucleare (INFN), Sez. di Napoli, Via Cinthia 9, I-80126 Napoli, Italy.}
\affiliation{Gran Sasso Science Institute, Via F. Crispi 7, I-67100, L' Aquila, Italy.}

\author{Rocco D'Agostino}
\email{rocco.dagostino@roma2.infn.it}
\affiliation{Dipartimento di Fisica, Universit\`a degli Studi di Roma ``Tor Vergata'', Via della Ricerca Scientifica 1, I-00133, Roma, Italy.}
\affiliation{Istituto Nazionale di Fisica Nucleare (INFN), Sez. di Roma ``Tor Vergata'', Via della Ricerca Scientifica 1, I-00133, Roma, Italy.}

\author{Orlando Luongo}	
\email{orlando.luongo@lnf.infn.it}
\affiliation{Istituto Nazionale di Fisica Nucleare, Laboratori Nazionali di Frascati, 00044 Frascati, Italy.}
\affiliation{School of Science and Technology, University of Camerino, I-62032, Camerino, Italy.}
\affiliation{Department of Mathematics and Applied Mathematics, University of Cape Town, Rondebosch 7701,
Cape Town, South Africa.}
\affiliation{Astrophysics, Cosmology and Gravity Centre (ACGC), University of Cape Town, Rondebosch 7701,
Cape Town, South Africa.}

\begin{abstract}
The limits of standard cosmography are here revised addressing the problem of error propagation during statistical analyses. To do so, we propose the use of  Chebyshev polynomials to parameterize cosmic distances. In particular, we demonstrate that building up rational Chebyshev polynomials significantly reduces error propagations with respect to standard Taylor series. This technique provides unbiased estimations of the cosmographic parameters and performs significatively better than previous numerical approximations. To figure this out, we compare rational Chebyshev polynomials with Pad\'e series. In addition, we theoretically evaluate the convergence radius of (1,1) Chebyshev rational polynomial and we compare it with the convergence radii of Taylor and Pad\'e approximations. We thus focus on regions in which convergence of Chebyshev rational functions is better than standard approaches. With this recipe, as high-redshift data are employed, rational Chebyshev polynomials remain highly stable and  enable one to derive highly accurate analytical approximations of Hubble's rate in terms of the cosmographic series. Finally, we check our theoretical predictions by setting bounds on cosmographic parameters through Monte Carlo integration techniques, based on the Metropolis-Hastings algorithm. We apply our technique to high-redshift cosmic data, using the JLA supernovae sample and the most recent versions of Hubble parameter and baryon acoustic oscillation measurements.  We find that cosmography with Taylor series fails to be predictive with the aforementioned data sets, while turns out to be much more stable using the Chebyshev approach.
\end{abstract}

\maketitle

\section{Introduction}

The cosmic acceleration is today confirmed by a large number of observations \cite{Riess98,Perlmutter99} and represents a consolidate challenge of modern cosmology. 
To disclose the physics behind it, model-independent techniques have been widely investigated during last years. Strategies toward model-independent treatments have as main target the determination of universe's expansion history without the need of postulating \emph{a priori} dark energy contributions. A particular attention is currently given to \emph{cosmography} \cite{Visser04,Cattoen07,Saini00,Cai11,Guimaraes11,Arabsalmani11,Carvalho11,Luongo11,Capozziello11}. Standard cosmography lies on Taylor expansions of cosmic distances. The method provides a powerful tool to study the dark energy evolution without assuming its functional form in the Hubble rate. Moreover, fixing limits over free cosmographic parameters alleviates degeneracy among models and enables to understand which paradigms are effectively favored directly with respect to data surveys. Although cosmography candidates as a robust tool to understand whether dark energy evolves or not, cosmic data unfortunately span on intervals $z\ge1$. This limit poses severe restrictions on cosmography and makes inapplicable Taylor expansions built up around $z\simeq0$.

One stratagem to overcome this problem is to parameterize Taylor expansions in terms of \emph{auxiliary variables}. Unfortunately, even this case turns out to be jeopardized by severe error propagations over the final outcomes. More recently, a further effort has been the use of Pad\'e approximation built up to converge at higher redshift domains \cite{Gruber14,Wei14}. In this case, however, the expansion orders are not fixed \emph{a priori} and this causes difficulties on evaluating the rate of convergence as $z\rightarrow\infty$. Thus, the limits of standard Taylor approach lying on $z\geq1$ are essentially alleviated but not fully fixed.

Motivated by the need of reducing relative uncertainties in cosmography, we here propose a new cosmographic technique based on Chebyshev polynomials. Chebyshev polynomials represent sequences of orthogonal polynomials, recursively-defined through trigonometric functions. In our approach we develop a new \emph{Chebyshev cosmography} adopting the strategy of building up rational approximations made by these polynomial functions. We demonstrate that, under the hypothesis of rational Chebyshev polynomials, distinguishing Chebyshev functions of first and second kinds is not relevant since the final output gives analogous results in both the cases. For simplicity, we limit our analysis on first kind Chebyshev polynomials only and we write the sequence of Chebyshev rational functions which better approximate Taylor series up to a certain order. We thus show that our Chebyshev ratios provide nodes in polynomial interpolation, minimizing cosmographic uncertainties leading to the most likely well-motivated approximation to cosmic distances. We even present theoretical motivations behind our choice by computing the convergence radii for different choices of polynomial approximations.

To check how well our model works we also study Pad\'e expansions and we compare the Chebyshev technique with them. We finally show the advantages of our procedure with respect to the old approaches using data surveys, confronting the cosmological quantities built from our method with the observables of the latest cosmological data sets. We adopt a Monte Carlo analysis employing the Metropolis-Hastings algorithm, choosing JLA supernova, baryon acoustic oscillation and differential age measurements. We show the goodness of our procedures comparing the outcomes coming from standard cosmography and our method, showing that error uncertainties are effectively reduced.

The structure of the paper is as follows.
In \Cref{sec:cosmography}, we review the general aspects of cosmography.
In \Cref{sec:Chebyshev}, we describe the mathematical features of the Chebyshev polynomials and present the method of the rational Chebyshev approximations.
In \Cref{sec:cosmography Chebyshev}, we derive a new model-independent formula for the luminosity distance, and compare the new method with the standard cosmographic procedures to verify the goodness of our approach.
In \Cref{sec:constraints}, we place observational limits on the cosmographic parameters through a confront with the most recent experimental data.
Finally, in \Cref{sec:conclusions} we summarize our findings and conclude.

\section{The cosmographic approach}
\label{sec:cosmography}

The study of the cosmic evolution can be done independently of energy densities by means of cosmography.
This model-independent technique only relies on the observationally justifiable assumptions of homogeneity and isotropy \cite{Weinberg72,Visser97,Harrison76}.
The great advantage of this method is that it allows one to reconstruct the dynamical evolution of the dark energy term without assuming any particular cosmological model.
Cosmography involves Taylor expansions of observable quantities that may, in principle, go up to any order. These expansions can be compared directly with data. The outcomes of this procedure ensure the independence from any postulated equation of state governing the evolution of the universe and, thus, help to break the degeneracy among cosmological models.

The homogenous and isotropic universe is governed by the single degree of freedom offered by $a(t)$, as demanded by the cosmological principle. Hence, following the most recent measurements of \cite{Planck15} and assuming a spatially flat universe\footnote{The assumption of flatness overcomes problems of degeneracy among the cosmographic parameters entering the expression of the luminosity distance \cite{Dunsby16}.}, we can expand $a(t)$ in Taylor series around present time $t_0$ \cite{Visser05,Visser15}:
\begin{equation}
a(t)=1+\sum_{k=1}^{\infty}\dfrac{1}{k!}\dfrac{d^k a}{dt^k}\bigg | _{t=t_0}(t-t_0)^k\ ,
\label{eq:scale factor}
\end{equation}
where $a(t_0)=1$.

The expansion above defines the cosmographic series \cite{Poplawski06,Poplawski07,Cattoen08,Xu11,Aviles12}:
\begin{align}
&H\equiv \dfrac{1}{a}\dfrac{da}{dt} \ , \hspace{1cm} q\equiv -\dfrac{1}{aH^2}\dfrac{d^2a}{dt^2}\ ,  \label{eq:H&q} \\
&j \equiv \dfrac{1}{aH^3}\dfrac{d^3a}{dt^3} \ , \hspace{0.5cm}  s\equiv\dfrac{1}{aH^4}\dfrac{d^4a}{dt^4}\ ,    \label{eq:j&s}
\end{align}
known in the literature as \textit{Hubble}, \textit{deceleration}, \textit{jerk} and \textit{snap} parameters\footnote{In principle, one may go further in the expansion and consider higher order coefficients. We limit our study up to the \emph{snap}, since the next cosmographic parameters are poorly constrained by observations \cite{Aviles12}.}. These coefficients are used to describe the expansion history of the universe at late times.

Using the relation $z=a^{-1}-1$ and \Cref{eq:scale factor}, one finds the Taylor series expansion of the luminosity distance as function of the redshift \cite{Busti15,Demianski12,Piedipalumbo15,Demianski17}:
\begin{align}
d_L(z)&=\dfrac{c}{H_0}z\bigg[1+\dfrac{z}{2}(1-q_0) -\dfrac{z^2}{6}\left(1-q_0-3q_0^2+j_0\right)+ \nonumber \\
&+\dfrac{z^3}{24}\left(2-2q_0-15q_0^2-15q_0^3+5j_0+10q_0j_0+s_0\right) \nonumber \\
&+ \mathcal{O}(z^4)\bigg].
\label{eq:luminosity distance}
\end{align}
The above expression for the luminosity distance can be used to obtain limits on the cosmographic parameters and study the low-$z$ dynamics of the universe with no need of any \emph{a priori} assumed cosmological model. In fact, plugging \Cref{eq:luminosity distance} into the definition
\begin{equation}
H(z)=c\,\left[\dfrac{d}{dz}\left(\dfrac{d_L(z)}{(1+z)}\right)\right]^{-1}\,,
\label{eq:Hubble rate}
\end{equation}
one gets
\begin{subequations}
\begin{align}
H(z)&\simeq H_0\left[1+H^{(1)}z+H^{(2)}\dfrac{z^2}{2}+H^{(3)}\dfrac{z^3}{6}\right]\,,\\
H^{(1)}&=1+q_0\,,\\
H^{(2)}&=j_0-q_0^2\,,\\
H^{(3)}&=3q_0^2+3q_0^3-j_0(3+4q_0)-s_0\,,
\label{eq:Taylor H(z)}
\end{align}
\end{subequations}
which describes the expansion history of the late-time universe up to the snap parameter.

\subsection{The convergence problem}

The limits of the standard cosmographic approach, based on the Taylor approximations, emerge when cosmological data at high redshifts are used to get information on the evolution of the dark energy term. In fact, the Taylor series converges if $z<1$, so that any cosmographic analysis employing data beyond this limit is plagued by severe restrictions.
A way to extend the radius of convergence of the Taylor series to high-redshift domains is represented by the method of rational approximations, among which the Pad\'e polynomials represent a relevant example \cite{Baker96}. The Taylor expansion of a generic function $f(z)$ is $
f(z)=\sum_{i=0}^\infty c_i z^i$, where $c_i=f^{(i)}(0)/i!$, whereas one defines the $(n,m)$ Pad\'e approximant of $f(z)$ as the rational polynomial
\begin{equation}
P_{n,m}(z)=\dfrac{\displaystyle{\sum_{i=0}^{n}a_i z^i}}{1+\displaystyle{\sum_{j=1}^{m}b_j z^j}}\,.
\label{eq:def Pade}
\end{equation}
Since by construction one requires that $b_0=1$, we have:
\begin{equation}
f(z)-P_{n,m}(z)=\mathcal{O}(z^{n+m+1})\ .
\end{equation}
The coefficients $b_i$ in \Cref{eq:def Pade} are thus determined by solving the following homogeneous system of linear equations \cite{Litvinov93}:
\begin{equation}
\sum_{j=1}^m b_j\ c_{n+k+j}=-b_0\ c_{n+k}\ ,
\end{equation}
valid for $k=1,\hdots,m$. All coefficients $a_i$ in \Cref{eq:def Pade} may be computed using the formula
\begin{equation}
a_i=\sum_{k=0}^i b_{i-k}\ c_{k}\,.
\end{equation}
The technique of Pad\'e approximations has been recently investigated in the context of cosmography to handle the divergence problems at high-$z$ domains \cite{Gruber14,Wei14}. In the next section, we present the method of rational Chebyshev polynomials that we will use to obtain a new cosmographic expression for the luminosity distance.

\section{Rational Chebyshev polynomials}
\label{sec:Chebyshev}

The method we propose here aims to optimize the technique of rational polynomials and consists of approximating the luminosity distance with a ratio of Chebyshev polynomials. In fact, the Pad\'e approximants are built up from the Taylor approximation of $d_L(z)$ whose error bars, by construction, rapidly increase as the redshift departs from zero. Motivated by this issue, we exploit the Chebyshev polynomials. Such a choice aims at reducing the uncertainties on the estimate of the cosmographic parameters.

The Chebyshev polynomials\footnote{Throughout the text, we refer to the Chebyshev polynomials of the \emph{first kind} simply as Chebyshev polynomials.} $T_n(z)$ are defined through the identity
\begin{equation}
T_n(z)=\cos(n\theta)\,,
\end{equation}
where $\theta=\arccos(z)$ and $n\in\mathbb{N}_0$. They form an orthogonal set with respect to the weighting function $w(z)=(1-z^2)^{-1/2}$ in the domain $|z|\leq1$  \cite{Chebyshev}:
\begin{equation}
\int_{-1}^{1}T_n(z)T_m(z) w(z)=
\begin{cases}
\pi\ , & n=m=0 \vspace{0.2cm}\\
\dfrac{\pi}{2} \delta_{nm}\ , & \text{otherwise}
\end{cases}
\end{equation}
where $\delta_{nm}$ is the Kronecker delta. The Chebyshev polynomials are generated from the recurrence relation
\begin{equation}
T_{n+1}(z)=2zT_n(z)-T_{n-1}(z)\ .
\end{equation}
The explicit expressions of the first five polynomials\footnote{We here truncate our analysis to the fifth order, since additional contributions go beyond our treatment. In so doing, we arrive to analyse up to snap parameter $s_0$.} that we will employ to build the new expression for $d_L(z)$ read \cite{numerical}:
\begin{equation}
\begin{aligned}
&T_0(z) = 1\ , \\
&T_1(z) = z \ ,\\
&T_2(z) = 2z^2 - 1 \ , \\
&T_3(z) = 4z^3 - 3z \ ,\\
&T_4(z) = 8z^4 - 8z^2 + 1 \ .
\end{aligned}
\label{eq:T_k}
\end{equation}
It is possible to express the powers of $z$ in terms of the Chebyshev polynomials according to the formula \cite{Litvinov93}:
\begin{equation}
z^n=2^{1-n}\sum_{k=0}^{[n/2]}a_k\binom{n}{k}T_{n-2k}(z)\ ,
\end{equation}
for $n>0$. Here, $[n/2]$ is the integer part of $n/2$,  $a_k=1/2$ if $k=n/2$ and $a_k=1$ if $a_k\neq n/2$, and $\binom{n}{k}$ are the binomial coefficients.

Let $f(z)\in L_w^2$, being $L_w^2$ the Hilbert space of the square-integrable functions with respect to the measure $w^{-1}(z)\ dz$.
Suppose we know the truncated Taylor series of $f(z)$ around the point $z=0$, $g(z)$. It is possible to obtain the polynomial of degree $n$, $\sum_{k=0}^n c_k T_k$ ,which gives the best approximation of $f(z)$ in the interval $[-1,1]$ in $L_w^2$. Formally, the Chebyshev series expansion of $f(z)$ reads
\begin{equation}
f(z)=\sum_{k=0}^{\infty} c_k T_k(z)\ ,
\label{eq:f(z)}
\end{equation}
where
\begin{align}
\begin{cases}
c_0= \dfrac{1}{\pi}\displaystyle{\int_{-1}^{1}}g(z)\ T(z)\ w(z)\ dz \ , \vspace{0.2cm} \\
c_k= \dfrac{2}{\pi}\displaystyle\int_{-1}^{1}g(z)\ T(z)\ w(z)\ dz\ , \hspace{0.5cm} k>0\ . \label{eq:c_k}
\end{cases}
\end{align}
Hence, we define the $(n,m)$ rational Chebyshev approximant as
\begin{equation}
R_{n,m}(z)=\dfrac{\displaystyle{\sum_{i=0}^n}\ a_i T_i(z)}{\displaystyle{\sum_{j=0}^m}\ b_j T_j(z)}\ .
\label{eq:ratio}
\end{equation}
For $b_0\neq 0$, through a redefinition of the coefficients, we can recast \Cref{eq:ratio} in the form
\begin{equation}
R_{n,m}(z)=\dfrac{\displaystyle{\sum_{i=0}^n}\ a_i T_i(z)}{1+\displaystyle{\sum_{j=1}^m}\ b_j T_j(z)}\ .
\label{eq:rational Chebyshev}
\end{equation}
Applying a similar procedure used to obtain the Pad\'e approximants, one can calculate the unknown coefficients $a_k$ and $b_k$ by equating \Cref{eq:f(z)} and \Cref{eq:rational Chebyshev} up to the $(n+m)$-th Chebyshev polynomial:
\begin{equation}
\sum_{k=0}^{\infty} c_k T_k=\dfrac{\displaystyle{\sum_{i=0}^n}\ a_i T_i}{1+\displaystyle{\sum_{j=1}^m}\ b_j T_j}+\mathcal{O}(T_{n+m+1})\ .
\end{equation}
By doing so, one gets:
\begin{align}
&(1+b_1T_1+\hdots+b_mT_m)(c_0+c_1T_1+\hdots)=  \nonumber \\
& a_0+a_1T_1+\hdots +a_nT_n+\mathcal{O}(T_{n+m+1})\ .
\label{eq:prod coeff}
\end{align}
To calculate the products of Chebyshev polynomials that occur in the left hand side of \Cref{eq:prod coeff}, one can make use of the trigonometric identity
\begin{equation*}
\cos(n\theta)\cos(m\theta)=\dfrac{1}{2}\Big[\cos\big[(n+m)\theta\big]+\cos\big[(n-m)\theta\big]\Big],
\end{equation*}
which leads to the relation
\begin{equation}
T_n(z)T_m(z)=\dfrac{1}{2}\Big[T_{n+m}(z)+T_{|n-m|}(z)\Big] .
\label{eq:prod Chebyshev}
\end{equation}
Thus, equating the terms with the same degree of $T$'s yields $(n+m+1)$ equations for the $(n+m+1)$ unknowns in \Cref{eq:rational Chebyshev}.

In the next section, we apply the mathematical procedure we have presented above to find a very accurate model-independent expression for the luminosity distance. We also compare our method with the cosmographic approaches developed so far in the literature.

\section{The Chebyshev cosmography}
\label{sec:cosmography Chebyshev}

We are here interested in approximating the luminosity distance with rational Chebyshev polynomials.
First, we need to express $d_L(z)$ in terms of Chebyshev polynomials according to \Cref{eq:f(z)}. To do that, we calculate the coefficients $c_k$ in \Cref{eq:c_k} where, in our case, $g(z)$ is the Taylor expansion given in \Cref{eq:luminosity distance}. Hence, the fourth-order Chebyshev expansion of the luminosity distance can be expressed as
\begin{equation}
d_L(z)=\dfrac{c}{H_0}\sum_{n=0}^4 c_n T_n(z)\ ,
\label{eq:dL new}
\end{equation}
where the coefficients $c_n$ read:
\begin{equation*}
\begin{aligned}
&c_0=\dfrac{1}{64} \Big[18 + 5 j_0 (1 + 2 q_0) - 3 q_0 \big(6 + 5 q_0 (1 + q_0)\big) + s_0\Big] ,\\
&c_1=\dfrac{1}{8}\left(7 - j_0 + q_0 + 3 q_0^2\right) ,\\
&c_2=\dfrac{1}{48}\Big[14 + 5 j_0 (1 + 2 q_0) - q_0 \big(14 + 15 q_0 (1 + q_0)\big) + s_0\Big] , \\
&c_3=\dfrac{1}{24}\big(-1 - j_0 + q_0 (1 + 3 q_0)\big) , \\
&c_4=\dfrac{1}{192}\Big[2 + 5 j_0(1 + 2 q_0) - q_0\big(2 + 15 q_0(1 + q_0)\big) + s_0\big] .
\end{aligned}
\end{equation*}
Thus, one can construct the rational Chebyshev approximations of $d_L(z)$ as in \Cref{eq:rational Chebyshev} starting from \Cref{eq:dL new}. We report some explicit expressions in \Cref{sec:appendix 1}.
Polynomials of high degrees will lead to more accurate approximations, even though these are the ones characterized by more complicated analytical forms, of course.

\subsection{Calibrating Chebyshev polynomials with the concordance model}

\begin{figure}
\begin{center}
\includegraphics[width=3.5in]{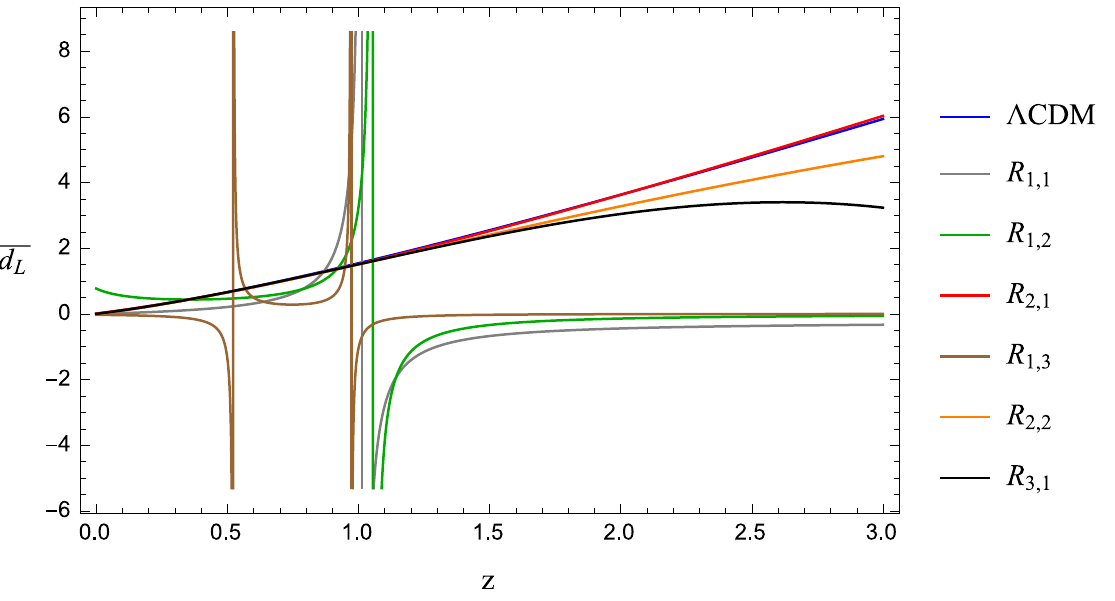}
\caption{Dimensionless luminosity distance as function of the redshift for rational Chebyshev approximations of the second ($R_{1,1}$), third ($R_{1,2}, R_{2,1}$) and fourth ($R_{1,3}, R_{2,2}, R_{3,1}$) degrees, compared to the $\Lambda$CDM model.}
\label{fig:dL_Rnm}
\end{center}
\end{figure}

To check the accuracy of various Chebyshev approximations, we compare them with the $\Lambda$CDM luminosity distance, $d_{L}(z)\Big|_{\Lambda\text{CDM}}$:
\begin{equation}
d_{L}(z)\Big|_{\Lambda\text{CDM}}=(1+z)\int_{0}^{z}\dfrac{c\ dz'}{H_{\Lambda\text{CDM}}(z')}\ ,
\label{eq:dL_LCDM}
\end{equation}
in which we have:
\begin{equation}
H_{\Lambda\text{CDM}}(z)= H_0\sqrt{\Omega_{m0}(1+z)^3+\Omega_{\Lambda}}\ .
\label{eq:H_LCDM}
\end{equation}
According to a spatially flat universe, we set $\Omega_{\Lambda}=1-\Omega_{m0}$, having $\Omega_{m0}$ the matter density at current time.  In the concordance paradigm case, the cosmographic parameters can be calculated in terms of $\Omega_{m0}$:
\begin{equation}
\begin{aligned}
&q_{0,\Lambda\text{CDM}}= -1+\dfrac{3}{2}\Omega_{m0}\ ,\\
&j_{0,\Lambda\text{CDM}}=1\ ,\\
&s_{0,\Lambda\text{CDM}}=1-\dfrac{9}{2}\Omega_{m0}\ .
\end{aligned}
\label{eq:cosmog param LCDM}
\end{equation}
As an indicative possibility, we fix $\Omega_{m0}=0.3$. So that from \Cref{eq:cosmog param LCDM} one gets:
\begin{equation}
\left\{
\begin{aligned}
&q_0=-0.55\ , \\
&j_0=1\ , \\
&s_0=-0.35\ .
\end{aligned}
\right.
\label{eq:indicat values}
\end{equation}
Using the values of \Cref{eq:indicat values}, in \Cref{fig:dL_Rnm} we show the behaviour with the redshift of $\overline{d_L}(z)\equiv\frac{H_0}{c}d_L(z)$ for different degrees of rational Chebyshev approximations.
In principle, to approximate the $\Lambda$CDM model some of the rational Chebyshev polynomials may present singularities turning out to give unsuitable outcomes.
To overcome this issue, the preferred rational approximations are those with $n-m\geq0$, in analogy to what happens for Pad\'e approximations \cite{Aviles14}. In particular, as practically checked the approximant $R_{2,1}(z)$ seems to give the most accurate approximation to the luminosity distance of the $\Lambda$CDM model. Assuming that the calibration with the concordance paradigm would be viable for any possible dark energy term, we assume that the most suitable approximation with Chebyshev polynomials comes from $R_{2,1}(z)$.

\begin{figure}
\begin{center}
\includegraphics[width=3.2in]{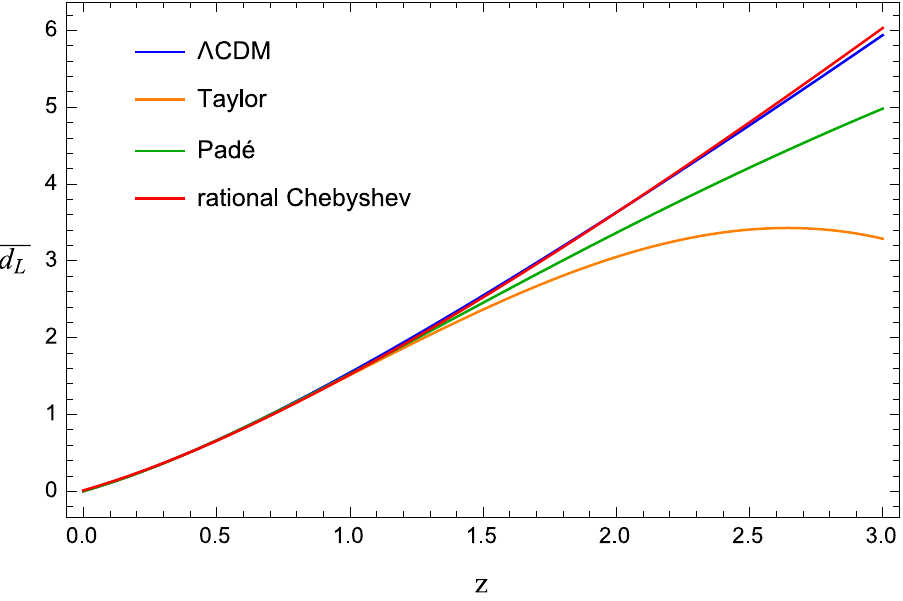}
\caption{Dimensionless luminosity distance as function of the redshift for the $\Lambda$CDM model and its fourth-order Taylor, (2,2) Pad\'e and (2,1) rational Chebyshev approximations.}
\label{fig:dL}
\end{center}
\end{figure}

To portray a qualitative representation of numerical improvements that one gains using our method, we compare $R_{2,1}(z)$ with the standard fourth-order Taylor expansion of $d_L(z)$ given in \Cref{eq:luminosity distance}, and with the (2,2) Pad\'e approximation of $d_L(z)$. We choose the (2,2) Pad\'e approximation since it has been argued that it is robustly characterized by good convergence properties \cite{Aviles14} as used in computational analyses.
We note that, while in the Taylor and Pad\'e approximations the \emph{snap} parameter shows up at the fourth order, in the rational Chebyshev polynomials it is present from the lowest degrees, since all the coefficients $c_k$ of \Cref{eq:dL new} have been calculated from the Taylor series expansion of $d_L(z)$ up to the \emph{snap} order (as confirmed in \Cref{eq:luminosity distance}).
For comparison, we report the expression of the (2,2) Pad\'e approximation of $d_L(z)$ in \Cref{sec:appendix 2}.
In \Cref{fig:dL}, we show the behaviour of $\overline{d_L}(z)$ for the various techniques. As can be seen, the Taylor approach fails when $z> 1$. Our Chebyshev cosmography stands out for the excellent approximation to the $\Lambda$CDM luminosity distance, resulting mostly more effective than Pad\'e approximations.

\subsection{The convergence radius}

We here argue how to \emph{broadcast} the above considerations to well-motivated theoretical scenarios. Thus, we wonder whether Chebyshev cosmography is expected to effectively improve the approximations to cosmic distances than standard cosmography. To do so, it behooves us to check how much the aforementioned approximations are stable to higher redshifts. Hence, one can test the ability of the various cosmographic techniques, able to describe high-redshift domains, by a direct comparison among the corresponding convergence radii, here defined by $\rho$.

\begin{figure}[h!]
\begin{center}
\includegraphics[width=3.2in]{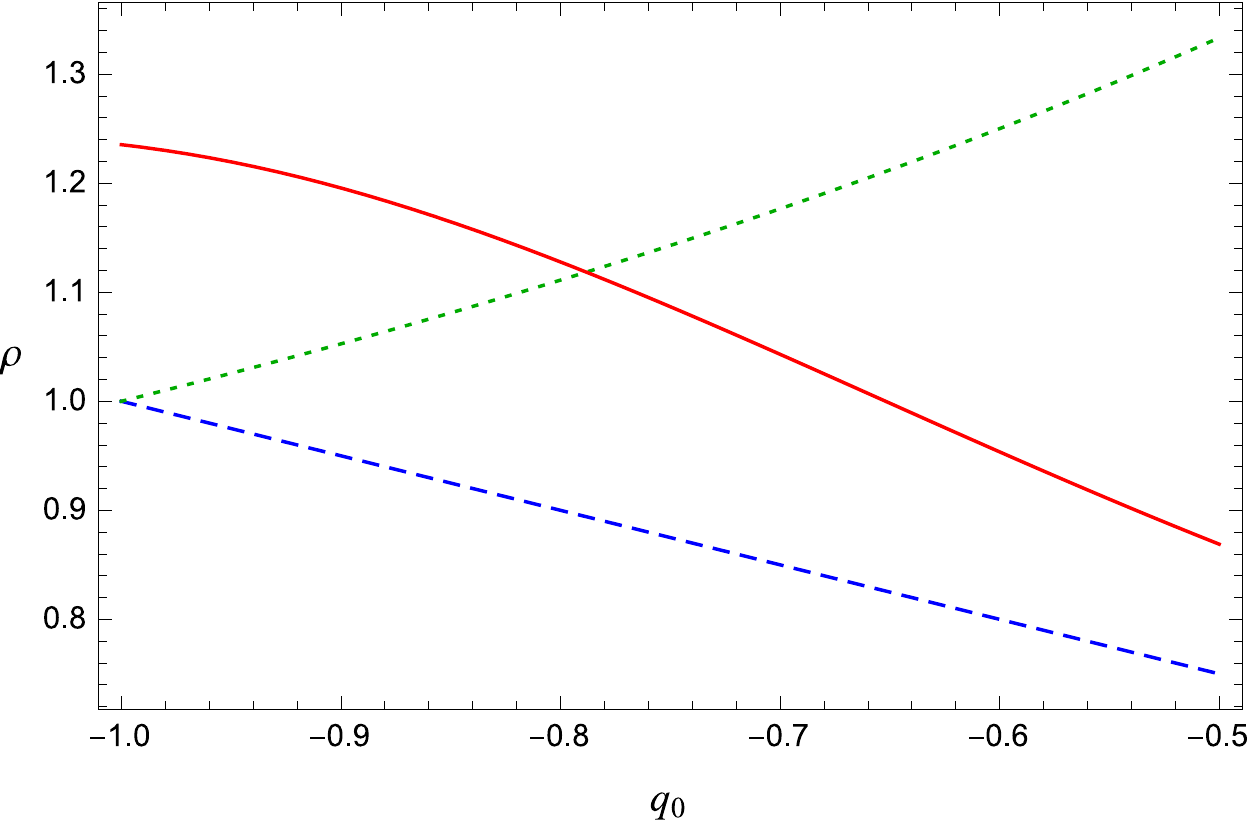}
\caption{Convergence radii for the second-order Taylor (dashed curve), (1,1) Pad\'e (dotted curve) and (1,1) rational Chebyshev (solid curve) approximations of the luminosity distance as a function of $q_0$. For the rational Chebyshev approximation we used the indicative values of $j_0=2$, $s_0=-1$.}
\label{fig:radii}
\end{center}
\end{figure}

As a simple example, we explicitly calculate the convergence radius of the (1,1) rational Chebyshev approximation of the luminosity distance, compared to the second-order Taylor series and to the (1,1) Pad\'e approximation. From \Cref{eq:rational Chebyshev,eq:T_k}, it holds
\begin{equation}
R_{1,1}(z)=\dfrac{a_0 T_0(z)+a_1T_1(z)}{1+b_1T_1(z)}=\dfrac{a_0+a_1 z}{1+b_1 z}\ ,
\label{eq:R_11}
\end{equation}
where the coefficients $\{a_0,a_1,b_1\}$ are expressed in terms of the cosmographic series as shown in \Cref{eq:R11}. We can rearrange \Cref{eq:R_11} as
\begin{equation}
R_{1,1}=\dfrac{a_0}{1+b_1 z}+\dfrac{a_1}{b_1}\left(1-\dfrac{1}{1+b_1 z}\right) ,
\end{equation}
and, after some algebra, one obtains
\begin{equation}
R_{1,1}=\dfrac{a_1}{b_1}+\left(a_0-\dfrac{a_1}{b_1}\right)\sum_{n=0}^{\infty}(-b_1)^n z^n\ .
\label{eq:new R_11}
\end{equation}
The geometric series in \Cref{eq:new R_11} converges for $|z|<1/|b_1|$, so that the convergence radius of the (1,1) rational Chebyshev approximation of $d_L(z)$ is
\begin{align}
\rho_{R_{1,1}}&=\dfrac{1}{|b_1|} \nonumber \\
&=\bigg|\dfrac{-3 (7 - j_0 + q_0 + 3 q_0^2)}{14 + 5 j_0 (1 + 2 q_0) - q_0 \big(14 + 15 q_0 (1 + q_0)\big) + s_0}\bigg|.
\end{align}
Analogous calculations show that the convergence radius of the (1,1) Pad\'e approximant for $d_L(z)$ is
\begin{equation}
\rho_{P_{1,1}}=\dfrac{2}{1-q_0}\ .
\end{equation}
The convergence radius of the second-order Taylor series of $d_L(z)$ is approximately given by
\begin{equation}
\rho_{d_{L,2}}=\dfrac{1-q_0}{2}\ .
\end{equation}
For the sake of completeness, the numerical values of $\rho_{R_{1,1}}, \rho_{P_{1,1}}$ and $\rho_{d_{L,2}}$ should be computed by using fitting results over the cosmographic coefficients. However, an analogous check can be made assuming the reference values \Cref{eq:cosmog param LCDM}. In such a case, one gets:
\begin{equation}
\left\{
\begin{aligned}
\rho_{R_{1,1}}&=1.014\ , \\
\rho_{P_{1,1}}&= 1.290\ ,\\
\rho_{d_{L,2}}&=0.775\ .
\end{aligned}
\right.
\end{equation}
These indicative results confirm the improvements of the rational polynomials in extending the radius of convergence with respect to the Taylor series. From the outcomes of \Cref{eq:cosmog param LCDM} we notice that the convergence radius of the Pad\'e approximation seems fairly better than the Chebyshev one. However, this is due to the choice made on the set $q_0,j_0$ and $s_0$. In \Cref{fig:radii} we plot the convergence radii for Taylor, Pad\'e and Chebyshev polynomials with a different set of cosmographic coefficients not calibrated over the concordance paradigm. In \Cref{fig:radii}, in particular, we show the regions in which the improvements of Chebyshev rational approximations become significant in terms of the convergence radius.  

\section{Observational constraints}
\label{sec:constraints}

In this section, we present the data we use to set bounds on the cosmographic parameters.

\subsection{Supernovae Ia}

In the present work, we test the Joint Light-curve Analysis (JLA) sample of 740 SNe of type Ia \cite{Betoule14} in the redshift interval $0.01<z<1.3$. All the SNe have been standardized using the SALT2 model \cite{Guy07} as fitter for their light curves.
The catalogue provides, for each SN, the redshift $z$, model-independent apparent magnitude in the $B$ band $(m_B)$, the stretch factor of the light-curve $(X_1)$, and the colour at maximum brightness $(C)$.
The theoretical distance modulus,
\begin{equation}
\mu_{th}(z)=25+5\log_{10}[d_L(z)]\ ,
\label{eq:mu}
\end{equation}
is parametrized as follows:
\begin{equation}
\mu_{obs}=m_B-(M_B-\alpha X_1+\beta C)\ ,
\label{eq:dist modulus}
\end{equation}
where the absolute magnitude is defined as
\begin{equation}
M_B=
\begin{cases}
 M, & \text{if}\ M_{host}<10^{10}M_{Sun}\\
  M+\Delta_M, & \text{otherwise}
 \end{cases}
\end{equation}
being $M_{host}$ the host stellar mass. The nuisance parameters \{$M$, $\Delta_M$, $\alpha$, $\beta$\} are fitted together with the cosmological parameters.
The normalized likelihood function of the SNe data is given by
\begin{equation}
\mathcal{L}_{SN}=\dfrac{1}{{|2\pi \textbf{C}|}^{1/2}}\exp\left[-\dfrac{1}{2}\left(\mu_{th}-\mu_{obs}\right)^\dagger \textbf{C}^{-1}\left(\mu_{th}-\mu_{obs}\right)\right],
\end{equation}
where $\textbf{C}$ is the $2220 \times 2200$ covariance matrix  constructed as in \cite{Betoule14}, which includes statistical and systematic uncertainties on the light-curve parameters.

\subsection{Observational Hubble Data}

The Hubble rate of a given cosmological model can be constrained by means of the model-independent measurements acquired through the differential age (DA) method, first presented in \cite{Jimenez02}. Such a technique uses red passively evolving galaxies as cosmic chronometers. In particular, one can obtain $H(z)$ by measuring the age difference of two close galaxies and using the relation:
\begin{equation}
H_{th}(z)=-\dfrac{1}{(1+z)}{\left(\dfrac{dt}{dz}\right)}^{-1}\ .
\end{equation}
The normalized likelihood function for the OHD data $(\mathcal L_{OHD})$ is built using a collection of 31 uncorrelated DA measurements of $H(z)$, which we report in \Cref{sec:appendix 3}:
\begin{equation}
\mathcal{L}_{OHD}=\dfrac{\exp\left[-\dfrac{1}{2}\displaystyle{\sum_{i=1}^{31}}\left(\dfrac{H_{th}(z_i)-H_{obs}(z_i)}{\sigma_{H,i}}\right)^2\right]} {\left[{(2\pi)}^{31}\displaystyle{\prod_{i=1}^{31}} \sigma_{H,i}^2\right]^{1/2}}\ .
\label{eq:likelihood_OHD}
\end{equation}

\subsection{Baryon Acoustic Oscillations}

Intensive studies on the large scale structures of the universe have been done thanks to galaxy surveys. The baryon acoustic oscillations that occur in the relativistic plasma come in the form of a characteristic peak in the galaxy correlation function.  The BAO measurements are usually given in the literature as $d_V^{th}(z)\equiv r_d/D_V(z)$, namely the ratio between the comoving sound horizon at the drag epoch $(r_d)$ and the spherically averaged distance measure introduced in \cite{Eisenstein05}:
\begin{equation}
D_V(z)= \left[\dfrac{{d_L}^2(z)}{(1+z)^2}\dfrac{c\ z}{H(z)}\right]^{1/3}.
\end{equation}
We construct the normalized likelihood function for the BAO data using the six uncorrelated and model-independent measurements given in \cite{Lukovic16}, which we list in \Cref{sec:appendix 3}:
\begin{equation}
\mathcal{L}_{BAO}=\frac{\exp{\left[-\dfrac{1}{2}\displaystyle{\sum_{i=1}^6}\left(\dfrac{d_V^{th}(z_i)-d_V^{obs}(z_i)}{\sigma_{d_{V,i}}}\right)^2\right]}}{\left[{(2\pi)}^6\displaystyle{\prod_{i=1}^6}\sigma_{d_{V,i}}^2\right]^{1/2}}\ .
\end{equation}

\subsection{Results of the Monte Carlo analysis}

\begin{figure*}
\begin{center}
\includegraphics[width=1\textwidth]{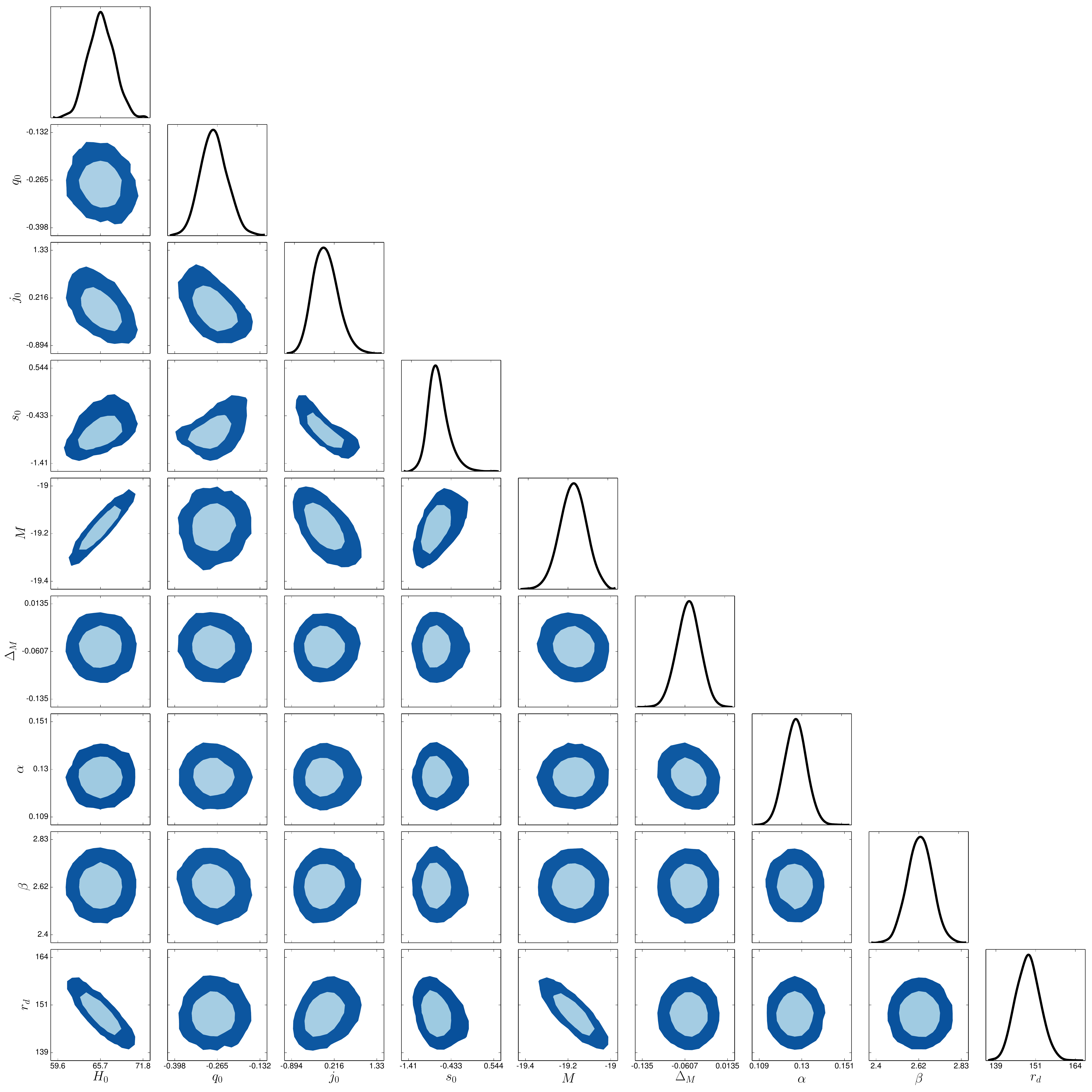}
\caption{68\% and 95\% confidence level contours and posterior distributions from the MCMC analysis of SN+OHD+BAO data for the fourth-order Taylor approximation of the luminosity distance. $H_0$ is expressed in Km/s/Mpc, and $r_d$ in Mpc.}
\label{fig:Taylor}
\end{center}
\end{figure*}

\begin{figure*}
\begin{center}
\includegraphics[width=1\textwidth]{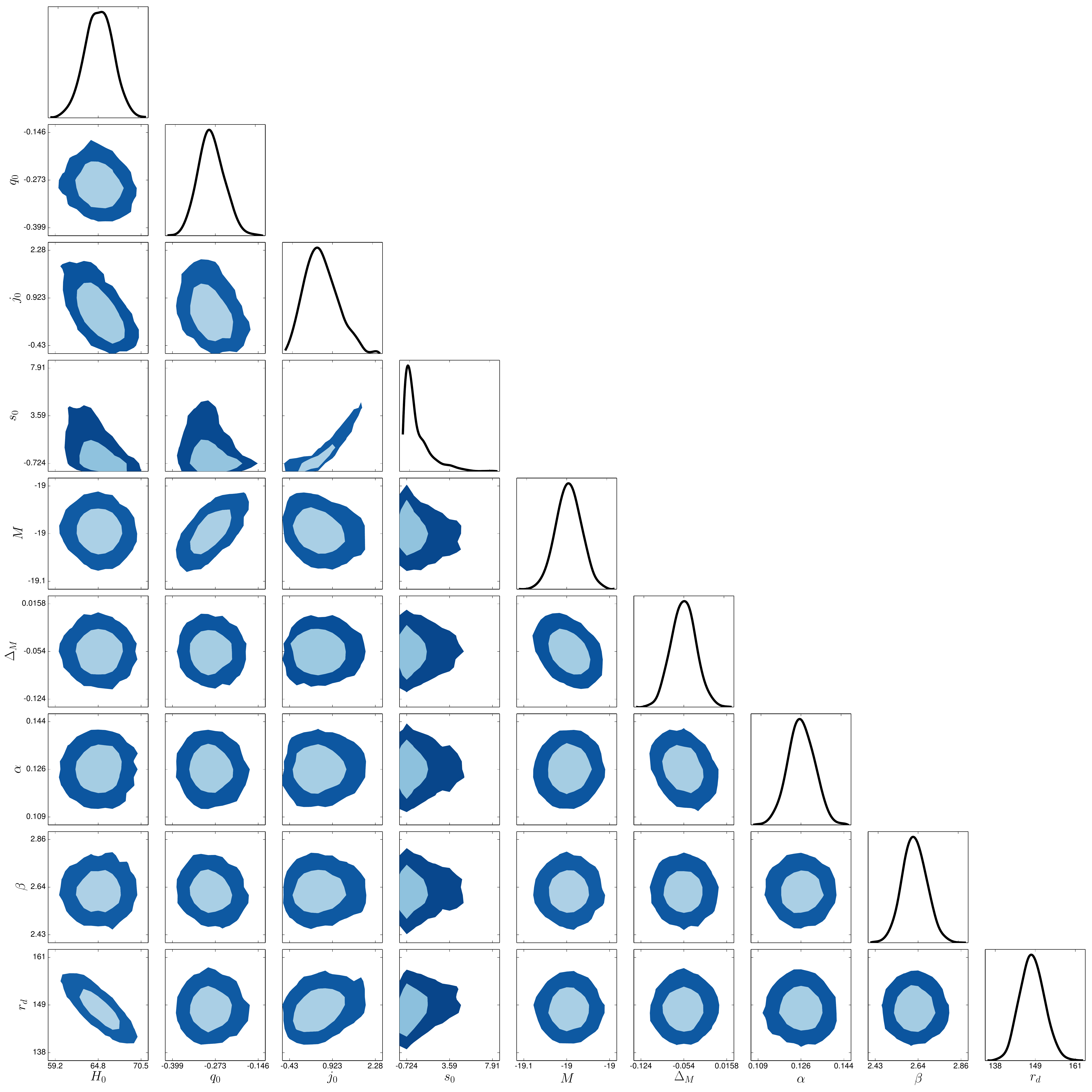}
\caption{68\% and 95\% confidence level contours and posterior distributions from the MCMC analysis of SN+OHD+BAO data for the (2,2) Pad\'e approximation of the luminosity distance. $H_0$ is expressed in Km/s/Mpc, and $r_d$ in Mpc.}
\label{fig:Pade}
\end{center}
\end{figure*}

\begin{figure*}
\begin{center}
\includegraphics[width=1\textwidth]{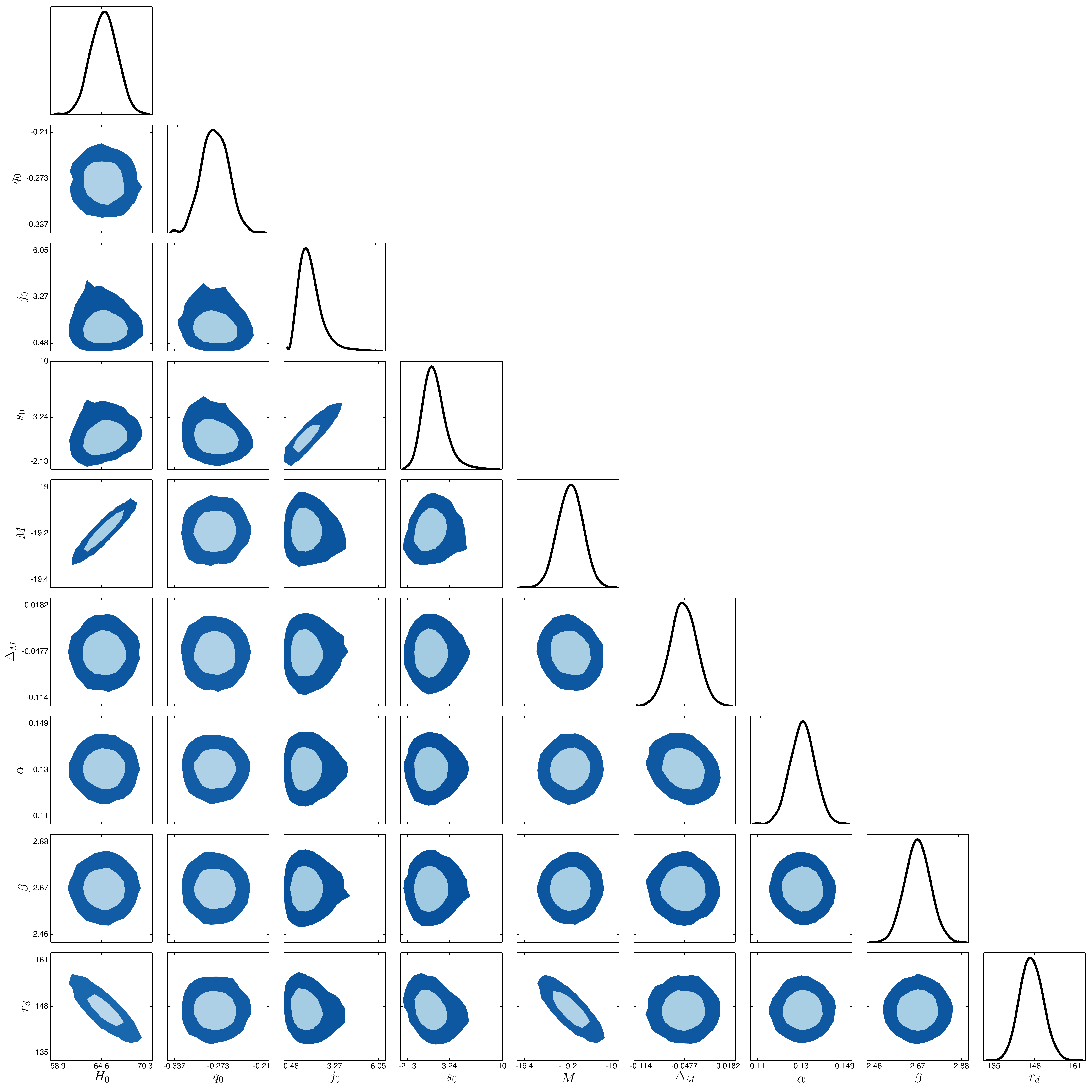}
\caption{68\% and 95\% confidence level contours and posterior distributions from the MCMC analysis of SN+OHD+BAO data for the (2,1) rational Chebyshev approximation of the luminosity distance. $H_0$ is expressed in Km/s/Mpc, and $r_d$ in Mpc.}
\label{fig:Chebyshev}
\end{center}
\end{figure*}

To test the different cosmographic approaches, we performed a Markov Chain Monte Carlo (MCMC) integration on the combined likelihood of the datasets we presented above:
\begin{equation}
\mathcal{L}_{joint}=\mathcal{L}_{SN}\times \mathcal{L}_{OHD}\times \mathcal{L}_{BAO}\ .
\end{equation}
We implemented the Metropolis-Hastings algorithm with the Monte Python code \cite{MontePython}, assuming uniform priors for the parameters (see \Cref{tab:priors}). The numerical results of our joint analysis are shown in \Cref{tab:joint fit}.
Also, in \Cref{fig:Taylor,fig:Pade,fig:Chebyshev}  we show the marginalized 2D $1\sigma$ and $2\sigma$ regions and the 1D posterior distributions for the cosmological and nuisance parameters in the case of the three cosmographic techniques.
Our results prove that the method of rational Chebyshev polynomials reduces the uncertainties in the estimate of the cosmographic parameters with respect to the other approaches, as shown by the relative errors in \Cref{tab:errors}.

An alternative approach is to start from the cosmographic expansion series of the Hubble rate and, then, evaluate the luminosity distance by numerical integrations, as pointed out in \cite{Aviles17}. However, it turns out that the analysis based on rational Chebyshev approximations of $H(z)$ does not lead to further reduction of the error propagation with respect to our original approach.

\begin{table}
\begin{center}
\renewcommand{\arraystretch}{1.5}
\begin{tabular}{c c  }
\hline
\hline
Parameters & Priors \\
\hline
$H_0$  & $(50,90)$ \\
$q_0$ & $(-10,10)$ \\
$j_0$ & $(-10,10)$\\
$s_0$ & $(-10,10)$\\
$M$ & $(-20,-18)$ \\
$\Delta_M$  & $(-1,1)$ \\
$\alpha$  & $(0,1)$ \\
$\beta$  & $(0,5)$ \\
$r_d$  & $(140,160)$ \\
\hline
\hline
\end{tabular}
\caption{Priors for parameters estimate in the MCMC numerical analysis. $H_0$ values are given in units of Km/s/Mpc, while $r_d$ values in units of Mpc.}
 \label{tab:priors}
\end{center}
\end{table}

\begin{table*}
\begin{center}
\setlength{\tabcolsep}{1em}
\renewcommand{\arraystretch}{1.8}
\begin{tabular}{c| c c c| c c c| c c c }
\hline
\hline
\multirow{2}{*}{Parameter} & \multicolumn{3}{c|}{Taylor} &  \multicolumn{3}{c|}{Pad\'e} &  \multicolumn{3}{c}{Rational Chebyshev} \vspace{-0.2cm} \\
& \footnotesize Mean & \footnotesize $1\sigma$ & \footnotesize$2\sigma$ & \footnotesize Mean & \footnotesize $1\sigma$ & \footnotesize $2\sigma$  & \footnotesize Mean & \footnotesize $1\sigma$ & \footnotesize $2\sigma$ \\
\hline
$H_0$ & $65.80$ & $^{+2.09}_{-2.11}$ & $^{+4.22}_{-4.00} $  & $64.94$ & $^{+2.11}_{-2.02}$ & $^{+4.12}_{-4.13}$ & $64.95$ & $^{+1.89}_{-1.94}$ & $^{+3.77}_{-3.77}$ \\
$q_0$ & $-0.276$ & $^{+0.043}_{-0.049}$ & $^{+0.093}_{-0.091}$ & $-0.285$ & $^{+0.040}_{-0.046}$ & $^{+0.087}_{-0.084}$  & $-0.278$ & $^{+0.021}_{-0.021}$ & $^{+0.041}_{-0.042}$ \\
$j_0$ & $-0.023$ & $^{+0.317}_{-0.397}$ & $^{+0.748}_{-0.685}$ & $0.545$ & $^{+0.463}_{-0.652}$ & $^{+1.135}_{-1.025}$ &  $1.585$ & $^{+0.497}_{-0.914}$ & $^{+1.594}_{-1.453}$ \\
$s_0$ & $-0.745$ & $^{+0.196}_{-0.284}$ & $^{+0.564}_{-0.487}$ & $0.118$ & $^{+0.451}_{-1.600}$ & $^{+3.422}_{-1.921}$ & $1.041$ & $^{+1.183}_{-1.784}$ & $^{+3.388}_{-3.087}$ \\
$M$ &  $-19.16$ & $^{+0.07}_{-0.07}$ & $^{+0.14}_{-0.14}$ & $-19.03$ & $^{+0.02}_{-0.02}$ & $^{+0.05}_{-0.05}$ & $-19.17$ & $^{+0.07}_{-0.07}$ & $^{+0.13}_{-0.13}$ \\
$\Delta_M$  & $-0.054$ & $^{+0.023}_{-0.022}$ & $^{+0.044}_{-0.045}$ & $-0.054$ & $^{+0.022}_{-0.023}$ & $^{+0.045}_{-0.045}$ & $-0.050$ & $^{+0.022}_{-0.022}$ & $^{+0.044}_{-0.045}$  \\
$\alpha$  & $0.127$ & $^{+0.006}_{-0.006}$ & $^{+0.012}_{-0.012}$ & $0.127$ & $^{+0.006}_{-0.006}$ & $^{+0.012}_{-0.012}$  & $0.130$ & $^{+0.006}_{-0.006}$ & $^{+0.012}_{-0.012}$  \\
$\beta$  & $2.624$ & $^{+0.071}_{-0.068}$ & $^{+0.136}_{-0.140}$ & $2.625$ & $^{+0.065}_{-0.069}$ & $^{+0.137}_{-0.135}$ &  $2.667$ & $^{+0.068}_{-0.069}$ & $^{+0.137}_{-0.135}$ \\
$r_d$ & $149.2$ & $^{+3.7}_{-4.1}$ & $^{+7.7}_{-7.5}$  & $148.6$ & $^{+3.5}_{-3.8}$ & $^{+7.5}_{-7.1}$ & $147.2$ & $^{+3.7}_{-4.0}$ & $^{+7.8}_{-7.5}$ \\
\hline
\hline
\end{tabular}
\caption{68\% and 95\% confidence level parameter constraints from the MCMC analysis of SN+OHD+BAO data for the fourth-order Taylor, (2,2) Pad\'{e} and (2,1) rational Chebyshev polynomial approximations of the luminosity distance. $H_0$ values are given in units of Km/s/Mpc, while $r_d$ values in units of Mpc.}
 \label{tab:joint fit}
\end{center}
\end{table*}

\begin{table*}
\begin{center}
\setlength{\tabcolsep}{1.2em}
\renewcommand{\arraystretch}{1.5}
\begin{tabular}{c| c c| c c| c c }
\hline
\hline
\multirow{2}{*}{Parameter} & \multicolumn{2}{c|}{Taylor}  & \multicolumn{2}{c|}{Pad\'e} & \multicolumn{2}{c}{Rational Chebyshev}  \vspace{-0.2cm} \\
&\footnotesize $1\sigma$ & \footnotesize $2\sigma$ &  \footnotesize $1\sigma$ &   \footnotesize $2\sigma$  &   \footnotesize $1\sigma$ &  \footnotesize $2\sigma$\\
\hline
$H_0$ & $3.19\%$ & $6.25\%$  & $3.17\%$ & $6.35\%$ & $2.95\%$ & $4.11\%$ \\
$q_0$ & $16.8\%$ & $33.5\%$ & $15.1\%$  & $ 30.1\%$ & $7.66\%$ & $14.8\%$ \\
$j_0$ & $1534\%$ & $3079\%$ & $102\%$ & $198\%$ & $44.5\%$ &	$96.1\%$ \\
$s_0$ & $32.2\%$ & $70.5\%$ & $866\%$ & $2258\%$ & $142\%$ & $311\%$	\\
\hline
\hline
\end{tabular}
\caption{68\% and 95\% relative uncertainties on the estimate of the cosmographic parameters from the MCMC analysis of SN+OHD+BAO data for the fourth-order Taylor, (2,2) Pad\'{e} and (2,1) rational Chebyshev polynomial approximations of the luminosity distance.}
 \label{tab:errors}
\end{center}
\end{table*}

\clearpage

\section{Summary and conclusions}
\label{sec:conclusions}

In this work, we revised the convergence problem in cosmography, adopting a new method to reduce error uncertainties and bias propagations as high-redshift data are used. We set bounds on the cosmographic series, considering rational approximations of the luminosity distance and we demonstrated that such approximations are also valid for any cosmic distances. In particular, our novel procedure is based on approximating the luminosity distance $d_L(z)$ with ratios of Chebyshev polynomials. Since, by definition, Chebyshev approximants are the most suitable polynomials in approximating functions, we expected to get fairly good outcomes in computing cosmographic series. Indeed, we found that our approach overcomes the convergence issues typical of standard cosmographic techniques based on Taylor approximations. This has been confirmed by computing convergence radii for different sets of cosmographic coefficients. We also compared our new approach with the consolidate procedure of Pad\'e expansions. We showed that both numerical bounds and convergence radii are improved under precise conditions. This naturally showed that Chebyshev rational polynomials are more suitable to describe the cosmic dynamics at $z>1$ than Pad\'e series. Bearing this in mind, through the predictions of the concordance $\Lambda$CDM model, we calibrated the orders of Chebyshev rational polynomials, providing the recipe of a new Chebyshev cosmography, which turned out to be more predictive than Taylor series at all redshift domains.  
We evaluated the (2,1) Chebyshev series, corresponding to a fourth-order Tayor series and to a (2,2) Pad\'e approximation. This showed that lower order Chebyshev series work better than higher ones constructed by Taylor and Pad\'e recipes.  We finally checked the goodness of our method, statistically combining the JLA supernova compilation, $H(z)$ differential age data and baryon acoustic oscillation measurements by performing Monte Carlo integrations based on the Metropolis-Hastings algorithm. To do so, we employed the free available Monte Python  code and we computed the corresponding contours up to the $95\%$ confidence level. The numerical improvements have been reported in terms of percentages on 1$\sigma$ and 2$\sigma$ confidence levels. The error percentages are severely lowered whereas the mean values are centred around intervals compatible with previous results on cosmography.

Our final outcomes forecast that the technique of Chebyshev cosmography substantially decreases the relative uncertainties on the estimates of cosmographic parameters respect to other previous approaches. This procedure candidates as a new way toward computing the cosmographic series and to better fix constraints on cosmography. Chebyshev cosmography is thus able to heal previous inconsistencies on convergence which plagued cosmography itself. Further, it enables to use high-redshift data surveys in cosmographic analyses with no large spreads over fitted coefficients. 

Future efforts will be devoted to match Pad\'e and Chebyshev techniques in different redshift domains. We will work on characterizing the cosmographic data over binning intervals in which one will be able to highly maximize both the convergence radii and mean values and to minimize both bias and uncertainties on free cosmographic coefficients. \\
\vspace{-1cm}

\section*{Acknowledgements}

This paper is based upon work from COST action CA15117  (CANTATA), supported by COST (European Cooperation in Science and Technology).
S.C. acknowledges the support of INFN (iniziativa specifica QGSKY).
R.D. thanks Federico Tosone for useful discussions on the Monte Python code.


\clearpage

\appendix

\begin{widetext}
 \section{Rational Chebyshev approximations of the luminosity distance}
 \label{sec:appendix 1}
 In this appendix, we write the rational Chebyshev approximations of the luminosity distance up to the fourth degree:
 \begin{align}
 R_{1,1}(z)&=-\dfrac{c}{H_0}\Big[(3 (2 - 2 q_0 - 15 q_0^2 - 15 q_0^3 + 5 j_0 (1 + 2 q_0) + s_0) +(1/(7 - j_0 + q_0 + 3 q_0^2))(-72 (7 - j_0 + q_0 + 3 q_0^2)^2 \nonumber \\
&+ 3 (18 + 5 j_0 (1 + 2 q_0) - 3 q_0 (6 + 5 q_0 (1 + q_0)) + s_0) (14 + 5 j_0 (1 + 2 q_0) - q_0 (14 + 15 q_0 (1 + q_0)) + s_0) \nonumber \\
&+2 (14 + 5 j_0 (1 + 2 q_0) - q_0 (14 + 15 q_0 (1 + q_0)) + s_0)^2) z)\Big/(576 (1 - ((14 + 5 j_0 (1 + 2 q_0) - q_0 (14 + 15 q_0 (1 + q_0)) \nonumber \\
&+ s_0) z)/( 3 (7 - j_0 + q_0 + 3 q_0^2))))\Big]\ ,
\label{eq:R11}
 \end{align}

 \begin{align}
 R_{1,2}(z)&=\dfrac{c}{H_0}(-((44184 + 5 j_0^3 (1 + 2 q_0) (1 + 10 q_0) (9 + 10 q_0) - 3 q_0 (32024 + q_0 (26948 + q_0 (4780+q_0 (-15938 +
      q_0 (2134 \nonumber \\
      &+ 15 q_0 (565 + 3 q_0 (249 + 25 q_0 (3 + q_0))))))))+7148 s_0 + q_0 (-5272 + q_0 (-16 + 3 q_0 (-32 + 3 q_0 (439 + 75 q_0 (2  \nonumber \\
      &+ q_0))))) s_0- 3 (-38 + q_0 (38 + 15 q_0 (1 + q_0))) s_0^2 + s_0^3+j_0 (49884 +  q_0 (34880 + q_0 (-41632 + q_0 (-9472  \nonumber \\
      &+ 135 q_0 (125 + q_0 (332 + 25 q_0 (5 + 2 q_0))))))+916 s_0 - 2 q_0 (-586 + 3 q_0 (439 + 75 q_0 (3 + 2 q_0))) s_0+15 (1 + 2 q_0) s_0^2) \nonumber \\
      &+j_0^2 (-222 + 59 s_0+ q_0 (8422 - 5 q_0 (17 + 15 q_0 (211 + 60 q_0 (2 + q_0)) - 60 s_0) + 300 s_0)))/(8 (-1 - j_0 + q_0 + 3 q_0^2) (7 \nonumber \\
   &- j_0 + q_0 + 3 q_0^2) +24 (7 - j_0 + q_0 + 3 q_0^2)^2 - (18 + 5 j_0 (1 + 2 q_0) - 3 q_0 (6 + 5 q_0 (1 + q_0)) + s_0)(14 + 5 j_0 (1 + 2 q_0) \nonumber \\
   & - q_0 (14 + 15 q_0 (1 + q_0)) s_0)))+4 (271 - 17 j_0 + 17 q_0 + 51 q_0^2+(4 (39106 - 56 j_0^3 + j_0^2 (1665 + q_0 (193 + 454 q_0))  \nonumber \\
   &+ 5 s_0 -  j_0 (14469 - 5 s_0+ q_0 (3492 + q_0 (10127 + q_0 (1033 + 1287 q_0)) + 5 s_0))+q_0 (14282 - 10 s_0 + q_0 (45365  \nonumber \\
        &- 10 s_0 +  q_0 (10501 + 3 q_0 (5082+q_0 (479 + 429 q_0)) + 15 s_0)))))/(-868 + j_0^2 (-7 + 100 q_0 (1 + q_0)) +  q_0 (-888 \nonumber \\
   &+ q_0 (-1412 + 3 q_0 (-64 + q_0 (139 + 75 q_0 (2 + q_0))))) +  32 s_0 - 2 q_0 (16 + 15 q_0 (1 + q_0)) s_0+s_0^2 + j_0 (544 + 10 s_0  \nonumber \\
   & - 2 q_0 (q_0 (139 + 75 q_0 (3 + 2 q_0)) - 2 (56+ 5 s_0))))) z)\Big/(576 (1 - (4 (214 - 5 j_0^2 (1 + 2 q_0) + j_0 (65 + 5 q_0 (33 + q_0 (8  \nonumber \\
   &+ 9 q_0)) - s_0) + 15 s_0 + q_0 (-204 - 5 q_0 (41+ 3 q_0 (18 + q_0 (4 + 3 q_0))) + s_0 + 3 q_0 s_0)) z)/(8 (-1 - j_0 + q_0 + 3 q_0^2) (7   \nonumber \\
   & -j_0 + q_0 + 3 q_0^2) + 24 (7 - j_0 + q_0 + 3 q_0^2)^2-(18 + 5 j_0 (1 + 2 q_0) - 3 q_0 (6 + 5 q_0 (1 + q_0)) + s_0) (14 + 5 j_0 (1 + 2 q_0) \nonumber \\
   & - q_0 (14 + 15 q_0 (1 + q_0)) + s_0)) -(4 (12 (-1-j_0 + q_0 + 3 q_0^2) (7 - j_0 + q_0 + 3 q_0^2) +  4 (1 + j_0 - q_0 (1 + 3 q_0))^2 \nonumber \\
   &- (14 + 5 j_0 (1 + 2 q_0) - q_0 (14 + 15 q_0 (1 + q_0)) + s_0)^2) (-1 + 2 z^2))/(3 (8 (-1 - j_0 + q_0 + 3 q_0^2) (7 - j_0 + q_0 + 3 q_0^2) \nonumber \\
   & + 24 (7 - j_0 + q_0 + 3 q_0^2)^2 - (18 + 5 j_0 (1 + 2 q_0) - 3 q_0 (6 + 5 q_0 (1 + q_0)) + s_0) (14 + 5 j_0 (1 + 2 q_0) - q_0 (14 \nonumber \\
   & + 15 q_0 (1 + q_0)) + s_0)))))\ ,
 \end{align}

 \begin{align}
 R_{2,1}(z)&=\dfrac{c}{H_0}(-((3 (16 (-1 - j_0 + q_0 + 3 q_0^2) (7 - j_0 + q_0 + 3 q_0^2) - (18 + 5 j_0 (1 + 2 q_0) - 3 q_0 (6 + 5 q_0 (1 + q_0)) + s_0) (14 \nonumber \\
  &+ 5 j_0 (1 + 2 q_0) - q_0 (14 + 15 q_0 (1 + q_0)) + s_0)))/(14 + 5 j_0 (1 + 2 q_0) - q_0 (14 + 15 q_0 (1 + q_0)) + s_0))+4 (47 - j_0 \nonumber \\
  &+ q_0 + 3 q_0^2 - (12 (-1 + q_0) (1 + j_0 - q_0 (1 + 3 q_0)))/(14 + 5 j_0 (1 + 2 q_0) - q_0 (14 + 15 q_0 (1 + q_0)) + s_0)) z \nonumber \\
  &-(4 (12 (-1- j_0 + q_0 + 3 q_0^2) (7 - j_0 + q_0 + 3 q_0^2) + 4 (1 + j_0 - q_0 (1 + 3 q_0))^2 - (14 + 5 j_0 (1 + 2 q_0) - q_0 (14  \nonumber \\
  &+ 15 q_0 (1 + q_0))+ s_0)^2) (-1 + 2 z^2))/(14 + 5 j_0 (1 + 2 q_0) - q_0 (14 + 15 q_0 (1 + q_0)) + s_0))\Big/(192 (1 + (4 (1 + j_0  \nonumber \\
  &- q_0 (1 + 3 q_0)) z)/(
     14+5 j_0 (1 + 2 q_0) - q_0 (14 + 15 q_0 (1 + q_0)) + s_0))) \ ,
 \end{align}

 \begin{align}
 R_{1,3}(z)&=-\dfrac{c}{H_0}\Big[(1561600 + 534784 s_0 + 41472 s_0^2 + 1344 s_0^3 + 143726992 z + 48790350 q_0^{10} z + 21262500 q_0^{11} z + 5315625 q_0^{12} z \nonumber \\
 &+15 j_0^4 (-489 - 840 q_0 + 69160 q_0^2 + 140000 q_0^3 + 70000 q_0^4) z - 14619744 s_0 z - 300520 s_0^2 z + 8232 s_0^3 z + 105 s_0^4 z  \nonumber \\
 &-56700 q_0^9 (48 + (-843 + 25 s_0) z)-675 q_0^8 (16576 + (92213 + 6300 s_0) z) - 180 q_0^7 (106896 + 7 (195524 + 6633 s_0) z) \nonumber \\
 &-8 q_0 (584896 + s_0 (118144 - 620452 z) - 44133176 z + 1029 s_0^3 z +  s_0^2 (5520 + 72838 z)) + 12 q_0^5 (677344 - 25916176 z \nonumber \\
 & + 23625 s_0^2 z + 6 s_0 (24080 + 153453 z))-4 q_0^2 (1575 s0^3 z + 12 s_0^2 (1372 + 9637 z) - 4 s_0 (-90320 + 653249 z) \nonumber \\
 &- 192 (-9337 + 762081 z))+2 q_0^4 (231903 s_0^2 z + 36 s_0 (27672 + 350209 z) + 4 (3013792 + 3194005 z))-4 q_0^3 (1575 s_0^3 z \nonumber \\
 & + 1008 s_0^2 (13 + 8 z) - 20 s_0 (3984 + 419287 z) - 32 (93610 + 3209821 z)) + 6 q_0^6 (23625 s0^2 z - 630 s_0 (-176 + 1205 z) \nonumber\\
 & - 4 (576840 + 17866679 z))-20 j_0^3 (-4592 + 878430 q_0^4 z + 787500 q_0^5 z + 315000 q_0^6 z + (-27002 + 63 s_0) z \nonumber \\
 &- 105 q_0^3 (512 + 5 (-277 + 40 s_0) z) - 2 q_0 (26144 + (374833 + 5187 s_0) z)-q_0^2 (100800 + (1004813 + 31500 s_0) z)) \nonumber \\
 & +2 j_0^2 (33693975 q_0^6 z + 21262500 q_0^7 z + 7087500 q_0^8 z +  5187 s_0^2 z + 348 (3456 + 52589 z) + s_0 (53600 + 751068 z) \nonumber \\
 & - 3150 q_0^5 (704 + 25 (-191 + 12 s_0) z)-30 q_0^3 (187152 + (2523232 + 67095 s_0) z) - 15 q_0^4 (380800 + (3084503 \nonumber \\
 &+ 126000 s_0) z) + 4 q_0 (587248 - 1747434 z + 7875 s_0^2 z + s_0 (53760 + 539493 z))+2 q_0^2 (15750 s_0^2 z + 105 s_0 (832  \nonumber \\
 &+ 539 z) - 4 (114020 + 7726509 z)))-4 j_0 (24395175 q_0^8 z + 12403125 q_0^9 z + 3543750 q_0^10 z - 525 s_0^3 z - 2 s_0^2 (2856 \nonumber \\
 & + 28939 z) - 4 s_0 (40160 + 41679 z) + 24 (-42472 + 1941285 z)-23625 q_0^7 (64 + 5 (-149 + 6 s_0) z) - 225 q_0^6 (22288  \nonumber \\
 &+ (150091+ 7875 s_0) z) - 30 q_0^5 (228144 + (2980493 + 92862 s_0) z)+q_0^3 (3597632 - 51966500 z + 70875 s_0^2 z  \nonumber \\
 &+ 48 s_0 (9590+  74537 z)) +q_0^2 (4962368 + 22362652 z + 77301 s_0^2 z + 8 s_0 (44604 + 585577 z))-2 q_0 (525 s_0^3 z  \nonumber \\
 &+21 s_0^2 (224 + 799 z) - 32 s_0 (-1111 + 59056 z) - 4 (65536 + 7419921 z))+q_0^4 (47250 s_0^2 z - 945 s_0 (-256 + 715 z)  \nonumber \\
 &- 2 (1501320+ 59053033 z))))\Big/(384 (-247152 + 68112 s_0 + 2340 s_0^2 + 50376 z - 59124 s_0 z - 3114 s_0^2 z - 63 s_0^3 z \nonumber \\
 &- 556608 z^2 - 66624 s_0 z^2 -  2448 s_0^2 z^2 + 502496 z^3 + 94928 s_0 z^3+2856 s_0^2 z^3 + 60 s_0^3 z^3 - 10125 q_0^9 z (-21 + 20 z^2)  \nonumber \\
 &- 2025 q_0^8 (-84 - 315 z + 80 z^2 + 300 z^3) - 135 q_0^7 (-2940 - 8955 z + 2800 z^2 + 8492 z^3)+27 q_0^6 (37440 - 25 (-1033 \nonumber \\
 &+ 63 s_0) z - 36608 z^2 + 300 (-83 + 5 s_0) z^3)+q_0^3 (-716352 - 1083156 z + 438528 z^2 + 53072 z^3 - 135 s_0^2 z (-21 + 20 z^2) \nonumber \\
 & - 144 s_0 (681 - 40 z - 676 z^2 + 32 z^3))+18 q_0^5 (61530 - 32087 z - 64232 z^2 + 15388 z^3 + 15 s_0 (-84 - 315 z + 8_0 z^2 \nonumber \\
 &+ 300 z^3))+3 j_0^3 (-820 - 1025 z + 976 z^2 + 1220 z^3 + 1000 q_0^3 z (-21 + 20 z^2) + 100 q_0^2 (-84 - 315 z + 80 z^2 + 300 z^3) \nonumber \\
 & + 10 q_0 (-840 - 1255 z + 800 z^2 + 1244 z^3))+q_0^2 (-1759344 + 571812 z + 933312 z^2 - 1082768 z^3 -  9 s_0^2 (-84 - 315 z \nonumber \\
 &+ 80 z^2 + 300 z^3) - 24 s_0 (2361 - 2725 z - 2596 z^2 + 1812 z^3))-2 q_0 (643992 - 369876 z - 448608 z^2 + 555088 z^3 \nonumber \\
 &+ 3 s_0^2 (-42 - 519 z + 40 z^2 + 476 z^3) + 4 s_0 (7668 - 13629 z - 7632 z^2 + 11188 z^3))+3 q_0^4 (238140 - 615186 z \nonumber \\
 &- 326640 z^2 + 428744 z^3 + 3 s_0 (-3360 - 14145 z + 3200 z^2 + 13252 z^3))+j_0^2 (-43692 - 90906 z + 6576 z^2 + 51944 z^3 \nonumber \\
 &- 13500 q_0^5 z (-21 + 20 z^2) - 1800 q_0^4 (-84 - 315 z + 80 z^2 + 300 z^3) + 3 s_0 (-840 - 1255 z + 800 z^2 + 1244 z^3) \nonumber \\
 &-75 q_0^3 (-2856 - 7917 z + 2720 z^2 + 7540 z^3) + 3 q_0^2 (116340 + (2825 - 6300 s_0) z - 116944 z^2 + 20 (-293 + 300 s_0) z^3) \nonumber \\
 &+2 q_0 (97470 - 169767 z - 106872 z^2 + 124508 z^3 + 30 s_0 (-84 - 315 z + 80 z^2 + 300 z^3)))+j_0 (692832 - 537540 z \nonumber \\
 &- 418176 z^2 + 631312 z^3 + 20250 q_0^7 z (-21 + 20 z^2) + 3375 q_0^6 (-84 - 315 z + 80 z^2 +300 z^3) + 3 s_0^2 (-84 - 315 z  \nonumber \\
 &+ 80 z^2+ 300 z^3)+4 s_0 (4158 - 9609 z - 4728 z^2 + 7268 z^3) + 180 q_0^5 (-2940 - 8955 z + 2800 z^2 + 8492 z^3) \nonumber \\
 &-9 q_0^4 (131820 - 225 (-229 + 28 s_0) z - 130224 z^2 + 100 (-509 + 60 s_0) z^3)+2 q0 (45 s0^2 z (-21 + 20 z^2) + 6 s_0 (4674 \nonumber \\
 & - 2755 z - 4744 z^2 + 2508 z^3) + 16 (18483 + 4656 z - 15180 z^2 + 12160 z^3))-6 q_0^3 (45 s_0 (-84 - 315 z + 80 z^2 + 300 z^3) \nonumber \\
 & + 8 (20580 - 20033 z - 21968 z^2 + 12932 z^3))-6 q_0^2 (s_0 (-3360 - 14145 z + 3200 z^2 + 13252 z^3) +  4 (17241 - 59711 z \nonumber \\
 &- 25508 z^2 + 43004 z^3)))))\Big]\ ,
 \end{align}

 \begin{align}
 R_{2,2}(z)&=-\dfrac{c}{H_0}\Big[(2808 + 3828 s_0 + 474 s_0^2 + 15 s_0^3 + 891712 z + 115776 s_0 z + 4560 s_0^2 z + 753744 z^2 + 142392 s_0 z^2 + 4284 s_0^2 z^2 \nonumber \\
  & +90 s_0^3 z^2 - 50625 q_0^9 (1 + 6 z^2) -10125 q_0^8 (15 - 16 z + 90 z^2) - 135 q_0^7 (1723 - 2800 z + 12738 z^2) +
 135 q_0^6 (-845 \nonumber \\
  &+ 10688 z - 7470 z^2 + 75 s_0 (1 + 6 z^2))-q_0^3 (-13140 + 440320 z - 79608 z^2 + 675 s_0^2 (1 + 6 z^2) +
    576 s_0 (2 + 275 z \nonumber \\
  &+ 12 z^2)) + 18 q_0^5 (4463 + 117160 z + 23082 z^2 + 75 s_0 (15 - 16 z + 90 z^2))+5 j_0^3 (183 - 688 z + 1098 z^2 + 3000 q_0^3 (1 \nonumber \\
  &+ 6 z^2) +  300 q_0^2 (15 - 16 z + 90 z^2) + 6 q_0 (311 - 800 z + 1866 z^2))-3 q_0^2 (9420 + 555072 z + 541384 z^2 +
    15 s_0^2 (15 - \nonumber \\
  &16 z + 90 z^2) + 8 s_0 (385 + 5300 z + 2718 z^2)) + 9 q_0^4 (15158 + 229520 z + 214372 z^2 + s_0 (2513 - 3200 z + 19878 z^2)) \nonumber \\
  &-6 q_0 (s_0^2 (79 - 40 z + 714 z^2) + 4 s_0 (319 + 4624 z + 5594 z^2) + 4 (351 + 61768 z + 69386 z^2))-3 j_0^2 (s_0 (-311 + 800 z \nonumber \\
  & - 1866 z^2) + 22500 q_0^5 (1 + 6 z^2) + 3000 q_0^4 (15 - 16 z + 90 z^2) + 125 q_0^3 (313 - 544 z + 2262 z^2) - 2 (1079 + 6040 z \nonumber \\
  &+ 12986 z^2)-5 q_0^2 (-293 + 37264 z - 1758 z^2 + 300 s_0 (1 + 6 z^2)) - 2 q_0 (6181 + 71720 z + 62254 z^2 + 50 s_0 (15 - 16 z \nonumber \\
  & + 90 z^2)))+3 j_0 (33750 q_0^7 (1 + 6 z^2) + 5625 q_0^6 (15 - 16 z + 90 z^2) + 5 s_0^2 (15 - 16 z + 90 z^2) + 4 s_0 (363 + 3400 z \nonumber \\
  &+ 3634 z^2) + 60 q_0^5 (1723 - 2800 z + 12738 z^2)+4 (1019 + 62496 z + 78914 z^2) - 15 q_0^4 (-1745 + 39664 z - 15270 z^2 \nonumber \\
  &+ 300 s_0 (1 + 6 z^2))+2 q_0 (75 s_0^2 (1 + 6 z^2) + 64 (36 + 1895 z + 1520 z^2) + 2 s_0 (427 + 8200 z + 3762 z^2)) -
 2 q_0^3 (225 s_0 (15 \nonumber \\
  &- 16 z + 90 z^2) + 8 (2477 + 42040 z + 19398 z^2))-2 q_0^2 (s_0 (2513 - 3200 z + 19878 z^2) + 12 (1433 + 18780 z \nonumber \\
  & + 21502 z^2))))\Big/(384 (-2348 - 324 s0 - 15 s_0^2 - 784 z - 200 s_0 z + 104 z^2 + 136 s_0 z^2 + 10 s_0^2 z^2 + 1125 q0^6 (-3 + 2 z^2) \nonumber \\
  &+ 90 q_0^5 (-75 - 4 z + 50 z^2) + q_0^4 (-6795 - 480 z + 3138 z^2)-6 q_0^3 (25 s_0 (-3 + 2 z^2) + 8 (-20 - 35 z + 16 z^2)) \nonumber \\
  & - 4 q_0 (-982 - 296 z + 52 z^2 + s_0 (-81 - 2 z + 34 z^2))+j_0^2 (-215 - 40 z + 122 z^2 + 500 q_0^2 (-3 + 2 z^2) + 20 q_0 (-75 \nonumber \\
  &- 4 z + 50 z^2)) + q_0^2 (8 (578 + 475 z - 146 z^2) - 6 s_0 (-75- 4 z + 50 z^2))-2 j_0 (1034 + 700 z - 212 z^2 + s_0 (75 + 4 z  \nonumber \\
  &-50 z^2) + 750 q_0^4 (-3 + 2 z^2) + 45 q_0^3 (-75 - 4 z + 50 z^2) + q_0^2 (-2265 - 160 z + 1046 z^2)+q_0 (970 + 780 z - 468 z^2 \nonumber \\
  & - 50 s_0 (-3 + 2   z^2)))))\Big]\ ,
 \end{align}

 \begin{align}
R_{3,1}(z)&= \dfrac{c}{H_0}(-((24 (-(((-1 - j_0 + q_0 (1 + 3 q_0)) (18 + 5 j_0 (1 + 2 q_0) - 3 q_0 (6 + 5 q_0 (1 + q_0)) + s_0))/1536) + ((7 - j_0 + q_0 \nonumber \\
 &+ 3 q_0^2) (2 + 5 j_0 (1 + 2 q_0) -   q_0 (2 + 15 q_0 (1 + q_0)) + s_0))/1536))/(-1 - j_0 + q_0 (1 + 3 q_0)))-(1/(-1 - j_0 + q_0 (1 \nonumber \\
 &+ 3 q_0))) 24 (-(1/192) (7 - j_0 + q_0 + 3 q_0^2) (-1 - j_0 + q_0 (1 + 3 q_0)) + ((18 + 5 j_0 (1 + 2 q_0) - 3 q_0 (6 + 5 q_0 (1 + q_0)) \nonumber \\
 &+ s_0) (2 + 5 j_0 (1 + 2 q_0) - q_0 (2 + 15 q_0 (1 + q_0)) + s_0))/6144+(((2 + 5 j_0 (1 + 2 q_0) - q_0 (2 + 15 q_0 (1 + q_0)) + s_0) (14 \nonumber \\
 &+  5 j_0 (1 + 2 q_0) - q_0 (14 + 15 q_0 (1 + q_0)) + s_0))/9216)) z-(1/(-1 - j_0 + q_0 (1 + 3 q_0))) 24 (((7 - j_0 + q_0 + 3 q_0^2) (2 \nonumber \\
 &+ 5 j_0 (1 + 2 q_0) - q_0 (2 + 15 q_0 (1 + q_0)) + s_0))/1536+(((-1 - j_0 + q_0 (1 + 3 q_0)) (2 + 5 j_0 (1 + 2 q_0) - q_0 (2 + 15 q_0 (1 \nonumber \\
 &+ q_0)) + s_0))/4608) - ((-1 - j_0 + q_0 (1 + 3 q_0)) (14 + 5 j_0 (1 + 2 q_0) - q_0 (14 + 15 q_0 (1 + q_0)) + s_0))/1152)(-1 + 2 z^2) \nonumber \\
 &- 1/(-1 - j_0 + q_0 (1 + 3 q_0))24 (-(1/576) (-1 - j_0 + q_0 (1 + 3 q_0))^2 + (2 + 5 j_0 (1 + 2 q_0) -  q_0 (2 + 15 q_0 (1 + q_0)) \nonumber \\
 &+ s_0)^2/36864+(((2 + 5 j_0 (1 + 2 q_0) - q_0 (2 + 15 q_0 (1 + q_0)) + s_0) (14 + 5 j_0 (1 + 2 q_0) - q_0 (14 + 15 q_0 (1 + q_0)) \nonumber \\
 & + s_0))/9216)) (-3 z + 4 z^3))\Big/(1 - ((2 + 5 j_0 (1 + 2 q_0) - q_0 (2 + 15 q_0 (1 + q_0)) + s_0) z)/( 4 (-1 - j_0 + q_0 (1 + 3 q_0))))\ .
 \end{align}

  \section{(2,2) Pad\'e approximant of the luminosity distance}
 \label{sec:appendix 2}

We report here the (2,2) Pad\'e approximation of the luminosity distance:
 \begin{align}
 P_{2,2}(z)&=\dfrac{c}{H_0}(6 z (10 + 9 z - 6 q_0^3 z + s_0 z - 2 q_0^2 (3 + 7 z) - q_0 (16 + 19 z) +
     j_0 (4 + (9 + 6 q_0) z))\Big/(60 + 24 z + 6 s_0 z - 2 z^2 \nonumber \\	
 &	 + 4 j_0^2 z^2 - 9 q_0^4 z^2 - 3 s_0 z^2 + 6 q_0^3 z (-9 + 4 z) + q_0^2 (-36 - 114 z + 19 z^2)\ .
 \end{align}

\clearpage

\section{Experimental data}
\label{sec:appendix 3}
Here, we list the compilations of OHD data and BAO data used to perform the Monte Carlo analysis.

\begin{table}[h]
\setlength{\tabcolsep}{2em}
\begin{center}
\begin{tabular}{c c c }
\hline
\hline
 $z$ &$H \pm \sigma_H$ &  Ref. \\
\hline
0.0708	& $69.00 \pm 19.68$ & \cite{Zhang14} \\
0.09	& $69.0 \pm 12.0$ & \cite{Jimenez02} \\
0.12	& $68.6 \pm 26.2$ & \cite{Zhang14} \\
0.17	& $83.0 \pm 8.0$ & \cite{Simon05} \\
0.179 & $75.0 \pm	4.0$ & \cite{Moresco12} \\
0.199 & $75.0	\pm 5.0$ & \cite{Moresco12} \\
0.20 &$72.9 \pm 29.6$ & \cite{Zhang14} \\
0.27	& $77.0 \pm 14.0$ & \cite{Simon05} \\
0.28	& $88.8 \pm 36.6$ & \cite{Zhang14} \\
0.35	& $82.1 \pm 4.85$ & \cite{Chuang12}\\
0.352 & $83.0	\pm 14.0$ & \cite{Moresco16} \\
0.3802	& $83.0 \pm 13.5$ & \cite{Moresco16}\\
0.4 & $95.0	\pm 17.0$ & \cite{Simon05} \\
0.4004	& $77.0 \pm 10.2$ & \cite{Moresco16} \\
0.4247	& $87.1 \pm 11.2$  & \cite{Moresco16} \\
0.4497 &	$92.8 \pm 12.9$ & \cite{Moresco16}\\
0.4783	 & $80.9 \pm 9.0$ & \cite{Moresco16} \\
0.48	& $97.0 \pm 62.0$ & \cite{Stern10} \\
0.593 & $104.0 \pm 13.0$ & \cite{Moresco12} \\
0.68	& $92.0 \pm 8.0$ & \cite{Moresco12} \\
0.781 & $105.0 \pm 12.0$ & \cite{Moresco12} \\
0.875 & $125.0 \pm 17.0 $ & \cite{Moresco12} \\
0.88	& $90.0 \pm 40.0$ & \cite{Stern10} \\
0.9 & $117.0 \pm 23.0$ & \cite{Simon05} \\
1.037 & $154.0 \pm 20.0$ & \cite{Moresco12} \\
1.3 & $168.0 \pm 17.0$ & \cite{Simon05} \\
1.363 & $160.0 \pm 33.6$ & \cite{Moresco15} \\
1.43	& $177.0 \pm18.0$ & \cite{Simon05} \\
1.53	& $140.0	\pm 14.0$ & \cite{Simon05} \\
1.75	 & $202.0 \pm 40.0$ & \cite{Simon05} \\
1.965& $186.5 \pm 50.4$ & \cite{Moresco15} \\
\hline
\hline
\end{tabular}
\caption{Differential age $H(z)$ data used in this work. The Hubble rate is given in units of km/s/Mpc.}
 \label{tab:OHD}
\end{center}
\vspace{-1cm}
\end{table}

\begin{table}[h]
\begin{center}
\setlength{\tabcolsep}{2em}
\begin{tabular}{ c c c c }
\hline
\hline
 $z$ &$d_V \pm \sigma_{d_V}$ &  Survey &  Ref. \\
\hline
0.106 & 0.336 $\pm$  0.015 &6dFGS   &  \cite{Beutler11}\\
0.15 & 0.2239 $\pm$ 0.0084 & SDSS DR7 & \cite{Ross15} \\
0.32 &  0.1181 $\pm$ 0.0023 & BOSS DR11 & \cite{Anderson14} \\
0.57 & 0.0726 $\pm$ 0.0007 & BOSS DR11 & \cite{Anderson14} \\
2.34 & 0.0320 $\pm$ 0.0016 & BOSS DR11 & \cite{Delubac15} \\
2.36 &  0.0329 $\pm$ 0.0012 & BOSS DR11 & \cite{Font-Ribera14}\\
\hline
\hline
\end{tabular}
\caption{BAO data used in this work.}
 \label{tab:BAO}
\end{center}
\end{table}

\end{widetext}

\end{document}